# TARGET SPEECH EXTRACTION BASED ON BLIND SOURCE SEPARATION AND X-VECTOR-BASED SPEAKER SELECTION TRAINED WITH DATA AUGMENTATION

*Zhaoyi Gu*[1], *Lele Liao*[1], *Kai Chen*[1], *Jing Lu*[1]

[1]Key Lab of Modern Acoustics, Institute of Acoustics, Nanjing University, Nanjing 210093, China

## ABSTRACT

Extracting the desired speech from a mixture is a meaningful and challenging task. The end-to-end DNN-based methods, though attractive, face the problem of generalization. In this paper, we explore a sequential approach for target speech extraction by combining blind source separation (BSS) with the x-vector based speaker recognition (SR) module. Two promising BSS methods based on source independence assumption, independent low-rank matrix analysis (ILRMA) and multi-channel variational autoencoder (MVAE), are utilized and compared. ILRMA employs nonnegative matrix factorization (NMF) to capture spectral structures of source signals and MVAE utilizes the strong modeling power of deep neural networks (DNN). However, the investigation of MVAE has been limited to the training with very few speakers and the speech signals of test speakers are usually included. We extend the training of MVAE using clean speech signals of 500 speakers to evaluate its generalization to unseen speakers. To improve the correct extraction rate, two data augmentation strategies are implemented to train the SR module. The performance of the proposed cascaded approach is investigated with test data constructed with real room impulse responses under varied environments.

## 1. INTRODUCTION

Target speech extraction aims at recovering speech signals of a desired speaker from an utterance deteriorated by interfering speech and noises given auxiliary information or assumptions about the target speaker. It is a more appropriate choice as an automatic speech recognition (ASR) front-end than blind source separation (BSS) [1] because it directly outputs the enhanced target speech whereas BSS outputs the separated signals of all existing sources.

Signal-processing based target speech extraction has been widely investigated, in which traditional methods such as beamforming and time-frequency masking are combined with BSS methods to obtain the source(s) of interest (SOI) [2–3]. Recently, two novel speech extraction algorithms based on a reformulated BSS mixing model are proposed to deal with the recovery of a non-Gaussian signal from a Gaussian background [4, 5]. The method is further extended to enhance the performance under a non-Gaussian background [6], and the extraction accuracy can be further improved by introducing a partially supervised structure with pilot signals in an online implementation [7, 8]. However, speech models used in these methods usually possess a fixed mathematical expression, which limits their capacity to characterize complex speech spectral structures.

Leveraging the advances in deep learning, DNN-based speech extraction has attracted increasing attention. In particular, several speaker aware neural networks have been proposed to guide an extraction network towards the learning of a target time-frequency mask by using an adaptation utterance of the target speaker [9–11]. However, as a pure supervised approach, end-to-end DNN-based methods face the challenge of generalization. Furthermore, most of the end-to-end methods focus on mono-channel processing, and are not easy to be extended to microphone arrays by effectively exploiting the spatial information.

Independent low-rank matrix analysis (ILRMA) [12] is a successful practice of introducing flexible source models into the framework of rule-base BSS algorithms. Recently, the proposed multi-channel variational autoencoder (MVAE) algorithm [13] incorporates a data-driven DNN-based signal model into the independent vector analysis (IVA) framework [14], and utilizes the conditional variational autoencoder (CVAE) [15] as the generative model of speech signals for each speaker. A fast MVAE algorithm [16] is further proposed to perform joint separation and classification of the source signals by training an auxiliary classifier VAE (ACVAE) under a closed-set setting. Although theoretically the MVAE effectively combines the benefits of both data-driven and rule-based methods, its success has only been validated with training on very few speakers, and its capability of generalizing to unseen speakers remains to be verified on larger datasets.

In this paper, we focus on the study of target speech extraction by sequentially combining ILRMA or MVAE with the x-vector based speaker recognition (SR) module [17]. Clean speech signals of 500 speakers are utilized in training the CVAE network to verify the generalization of MVAE to unseen speakers. In addition, we compare two different data augmentation strategies for the training of the SR module. Comprehensive discussions regarding extraction accuracy and signal-to-distortion ratio (SDR) are presented based on experiments using mixed signal with real room impulse responses.

## 2. MVAE BASED MULTI-CHANNEL BLIND SOURCE SEPARATION

### 2.1. Problem formulation

The signal model of a determined situation is assumed, where an array of $M$ microphones is utilized to capture the signals from $M$ sources. After transforming the signals into short-time Fourier transform (STFT) domain and ignoring the noise, the mixing model can be represented with an instantaneous model as

$$\mathbf{x}_{ft} = \mathbf{A}_f \mathbf{s}_{ft} \quad (1)$$

where $\mathbf{s}_{ft} = [s_{ft,1}, s_{ft,2}, \ldots, s_{ft,M}]^T$, $\mathbf{x}_{ft} = [x_{ft,1}, x_{ft,2}, \ldots, x_{ft,M}]^T$ are the multi-channel vectors containing the source and observed signals at frequency bin $f$ and time frame $t$ respectively, with $f = 1, \ldots, F$ and $t = 1, \ldots, T$. Subscript $m$ denotes the index of either sources or microphones based on the context, and $[\cdot]^T$ is a notation for non-conjugate transposition. $\mathbf{A}_f$ is an $M \times M$ complex-valued mixing matrix containing the information of the room impulse responses in the frequency domain. When $\mathbf{A}_f$ is invertible, estimated source signals $\mathbf{y}_{ft} = [y_{ft,1}, y_{ft,2}, \ldots, y_{ft,M}]^T$ can be obtained by multiplying a demixing matrix $\mathbf{W}_f = [\mathbf{w}_{f,1}, \mathbf{w}_{f,2}, \ldots, \mathbf{w}_{f,M}]^H$ as

$$\mathbf{y}_{ft} = \mathbf{W}_f \mathbf{x}_{ft} \quad (2)$$

where $[\cdot]^H$ is a notation for conjugate transposition.

Let $\mathbf{S}_m = \{s_{ft,m}\}_{ftm}$ be the vector containing the $m$th source signals at all $(t, f)$ bins, which is assumed in MVAE to follow a local Gaussian model (LGM)

$$p_{\mathbf{s}_m}(\mathbf{S}_m) = \prod_{ft} \mathcal{N}_c(s_{ft,m} | 0, v_{ft,m}) \quad (3)$$

where signals from different $(t, f)$ bins are independently characterized using a zero-mean circular symmetric Gaussian distribution with $v_{ft,m}$ being the estimated power spectral density. Assuming the independence among different sources, the signal model for the observed signals $\mathbf{x}_{ft}$ at time-frequency bin $(t, f)$ can be derived as

$$p_{\mathbf{x}_{ft}}(\mathbf{x}_{ft}) = |\det \mathbf{W}_f|^2 \prod_m \mathcal{N}_c(y_{ft,m} | 0, v_{ft,m}) \quad (4)$$

where $\det(\cdot)$ is the determinant notation. Hence, the log likelihood function of the demixing matrix $\mathbf{W}_f$ and the source model parameters $v_{ft,m}$ given the observed signals $\mathbf{x}_{ft}$ becomes

$$\mathcal{L}_{\mathbf{W},\mathbf{v}} \stackrel{c}{=} -\sum_{mft}\left(\log v_{ft,m} + \frac{|y_{ft,m}|^2}{v_{ft,m}}\right) + 2T\sum_f \log|\det \mathbf{W}_f| \quad (5)$$

where $\mathbf{v}$ denotes the whole set of $\{v_{ft,m}\}_{ftm}$. Dependencies among time-frequency bins in $v_{ft,m}$ given source $m$ are additionally characterized with a CVAE decoder model using a CNN architecture, so that theoretically the permutation ambiguity can be simultaneously eliminated during the separation process [14].

### 2.2. Conditional variational autoencoder

CVAE learns a multi-modal deep generative model of the speech spectrograms $\mathbf{S}_j$ by introducing a hierarchical structure with a latent vector $\mathbf{z}_j$, and a condition vector $\mathbf{c}_j$ that contains class information. Specifically, a disentangled representation between the latent and the speaker identity space is prompted by maximizing the following log likelihood function

$$\mathcal{L}_{\text{CVAE}} = \sum_j \log \int p_\psi(\mathbf{S}_j | \mathbf{z}_j, \mathbf{c}_j) p(\mathbf{z}_j) d\mathbf{z}_j \quad (6)$$

where $p_\psi(\mathbf{S}_j|\mathbf{z}_j, \mathbf{c}_j)$ is the conditional probability distribution of speech spectrogram and $p(\mathbf{z}_j)$ represents the prior distribution of the latent vector. Subscript $j$ is used instead of $m$ to denote an arbitrary CVAE training example. Since (6) cannot be directly optimized, a lower bound is derived using Jenson's inequality as

$$\begin{aligned}\mathcal{Q}_{\text{CVAE}} &= \sum_j \int q_\phi(\mathbf{z}_j|\mathbf{S}_j,\mathbf{c}_j) \log \frac{p_\psi(\mathbf{S}_j|\mathbf{z}_j,\mathbf{c}_j)p(\mathbf{z}_j)}{q_\phi(\mathbf{z}_j|\mathbf{S}_j,\mathbf{c}_j)} d\mathbf{z}_j \\ &\stackrel{c}{=} \sum_j \mathbb{E}_{q_\phi(\mathbf{z}_j|\mathbf{S}_j,\mathbf{c}_j)}\left[\log p_\psi(\mathbf{S}_j|\mathbf{z}_j,\mathbf{c}_j)\right] \\ &\quad - \sum_j \text{KL}\left(q_\phi(\mathbf{z}_j|\mathbf{S}_j,\mathbf{c}_j) \| p(\mathbf{z}_j)\right),\end{aligned} \quad (7)$$

where $\text{KL}(p\|q)$ denotes the Kullback-Leibler divergence between two probability density functions (PDF). To optimize (7), CVAE employs an encoder network to estimate the posterior distribution $q_\phi(\mathbf{z}_j|\mathbf{S}_j, \mathbf{c}_j)$ of the latent vector, and a decoder network to produce the generative distribution $p_\psi(\mathbf{S}_j|\mathbf{z}_j, \mathbf{c}_j)$, where $\phi, \psi$ are symbols denoting the parameters of the two networks respectively.

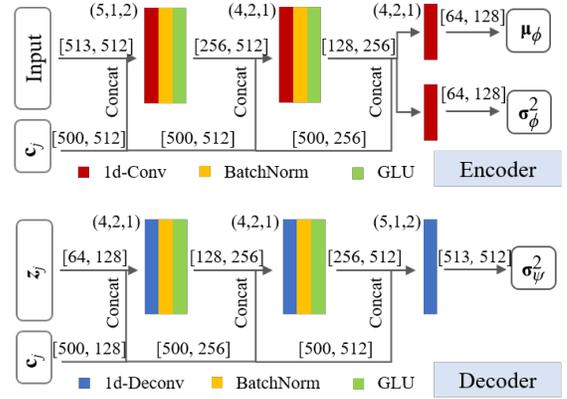

**Fig. 1**. Configurations of the CVAE network. The pair in square bracket denotes the channel dimension and width; the triplets in parenthesis denote the kernel size, stride and padding respectively.

In this work, $\mathbf{c}_j$ is assigned as a one-hot vector indicating the speaker identity, and $p_\psi(\mathbf{S}_j|\mathbf{z}_j, \mathbf{c}_j)$, $q_\phi(\mathbf{z}_j|\mathbf{S}_j, \mathbf{c}_j)$, $p(\mathbf{z})$ are assumed to be Gaussian distributions. Specifically,

$$p_\psi(\mathbf{S}_j|\mathbf{z}_j,\mathbf{c}_j) = \prod_{ft} \mathcal{N}_c\left(s_{ft,j} | 0, \sigma_\psi^2(f,t;\mathbf{z}_j,\mathbf{c}_j)\right) \quad (8)$$

$$q_\phi(\mathbf{z}_j|\mathbf{S}_j,\mathbf{c}_j) = \prod_d \mathcal{N}\left(z_{d,j} | \mu_\phi(d;\mathbf{S}_j,\mathbf{c}_j), \sigma_\phi^2(d;\mathbf{S}_j,\mathbf{c}_j)\right) \quad (9)$$

$$p(\mathbf{z}_j) = \mathcal{N}(\mathbf{z}_j | 0, \mathbf{I}) \quad (10)$$

where $\sigma_\psi^2(f, t; \mathbf{z}_j, \mathbf{c}_j)$ denotes the generated source spectrogram, and $\mu_\phi(d; \mathbf{S}_j, \mathbf{c}_j)$ and $\sigma_\phi^2(d; \mathbf{S}_j, \mathbf{c}_j)$ are the posterior mean and variance of the $d$th element in the latent vector. Note that (8) possesses the form of an LGM, so that the training objective of CVAE is in consistent with the MVAE objective depicted in (5). Fig. 1 shows the configuration of the CVAE network.

## 2.3. MVAE optimization process

After training the CVAE model, the decoder network is utilized as a deep source model in MVAE. Since the CVAE training utterances might be different from the source signals in terms of global energy, a scale factor $g_m$ is introduced to compensate for the difference. Therefore, the source model used in MVAE can be specified as

$$p_{\mathbf{s}_m}(\mathbf{S}_m) = \prod_{ft} \mathcal{N}_c\left(s_{ft,m} \big| 0, g_m \sigma_\psi^2(f,t;\mathbf{z}_m,\mathbf{c}_m)\right) \quad (11)$$

Incorporating (11) into the MVAE objective function (5), updating rules for $g_m$, $\mathbf{W}_f$, $\mathbf{z}_m$ and $\mathbf{c}_m$ can be derived as follows

$$g_m = \frac{1}{FT} \sum_{ft} \frac{|y_{ft,m}|^2}{\sigma_\psi^2(f,t;\mathbf{z}_m,\mathbf{c}_m)} \quad (12)$$

$$\mathbf{V}_{f,m} = \frac{1}{T} \sum_t \frac{\mathbf{x}_{ft}\mathbf{x}_{ft}^H}{g_m \sigma_\psi^2(f,t;\mathbf{z}_m,\mathbf{c}_m)} \quad (13)$$

$$\mathbf{w}_{f,m} = (\mathbf{W}_f \mathbf{V}_{f,m})^{-1} \mathbf{e}_m \quad (14)$$

$$\mathbf{w}_{f,m} = \mathbf{w}_{f,m} \big/ \sqrt{\mathbf{w}_{f,m}^H \mathbf{V}_{f,m} \mathbf{w}_{f,m}} \quad (15)$$

where (12) is derived by setting the derivative of (5) with respect to $g_m$ to zero and (13) – (15) are derived using the iterative-projection (IP) method [18]. $\mathbf{e}_m$ is the $m$th column of an $M \times M$ identity matrix, and $(\cdot)^{-1}$ is the notation for matrix inversion. Both $\mathbf{z}_m$ and $\mathbf{c}_m$ are updated via back propagation.

## 3. TARGET SPEECH EXTRACTION

After the multi-channel outputs are obtained using either ILRMA or MVAE as described in section 2, the target speech signals are identified and selected by matching the x-vectors of all channel candidates with that of the enrollment utterance.

### 3.1. x-vector based SR system

The x-vector embedding system used in the paper has the same network structure as illustrated in [19], and the embeddings are extracted from the affine component of layer segment 6. The input features are 30-dimensional MFCCs with a frame length of 64 ms and a frame hop of 16 ms. For each batch of data, the input length is randomly chosen within a range of 160 to 200 frames, and each training sample is mean-and-variance-normalized. Energy-based VAD is implemented beforehand in the time domain to filter out the silence segments in the training data. During testing, the input MFCCs are segmented with a window length of 180 frames and an overlap of 50%. The final embedding vector is computed as the average of x-vectors extracted from each window.

The x-vectors are centered and the dimensionality is reduced to 128 using linear discriminant analysis (LDA). After that, the embeddings are length normalized and a PLDA backend is trained using the Bob toolbox [20, 21]. The BSS output whose x-vector has the highest joint probability with the enrolled x-vector is extracted as the target signal.

### 3.2 Data augmentation

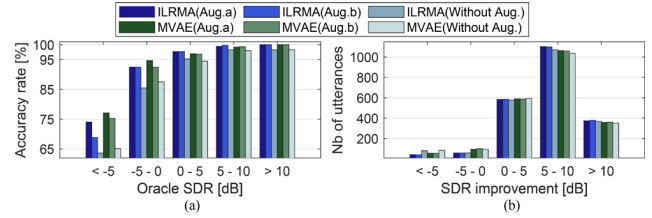

**Fig. 2**. Histogram of (a) extracted accuracy rate with respect to oracle SDR and (b) utterance distribution with respect to extracted SDR improvement respectively.

Usually the residual interference is inevitable in the output of the BSS, so training the x-vector system with augmented noisy data may lead to higher extraction rate. We explore two data augmentation strategies, denoted as Aug.a and Aug.b, as follows.

Aug.a employs the output signals of ILRMA algorithm given binaural mixed signals of 5 to 30 seconds generated by convolving the clean training utterances with simulated room impulse responses (RIRs). In our experiment, a total number of 4410 RIRs are created so as to cover a wide range of mixing scenario. Specifically, we consider seven different conditions with reverberation times (RT) ranging from 0.1 s to 0.7 s with a 0.1 s interval. Under each condition, 21 regular sized rooms are created, with the length, width and height lying in the range of [3 m, 15 m], [3 m, 15 m] and [2 m, 4 m] respectively. For each room, 30 RIRs are generated using the image model method [22] according to the following configurations. Microphone arrays with an average spacing of 0.2 m are randomly located in the room with a minimum distance of [1 m, 1.5 m] to the wall. Both sources lie on the same side of the array and the direction of arrival (DOA) interval is selected in the range of [20°, 160°]. The height of sources and microphones are set to be 1.2 m, and the distances between array center and sources are randomly chosen from [0.5 m, $r_c$ + 0.5 m], where $r_c$ denotes the critical distance. In total, two million training samples are generated by randomly picking utterances and offsets from the clean and artificial separated signals, and the number of samples extracted from the clean and augmented datasets are roughly the same.

Aug.b adopts a similar generic augmentation strategy as described in [19], where only augmented copies containing babble noise and reverberation are included in our implementation. The same RIR datasets [23] are used to involve reverberation and 3 to 7 utterances of different speakers with signal-to-noise-ratio (SNR) in the range of 13 – 20 dB are added to form the babble noise as implemented in the Kaldi SRE16 recipe [24]. For Aug. b, we generate one million training samples respectively for the clean dataset, the reverberated dataset and the clean signals deteriorated with babble noise.

## 4. EXPERIMENTAL SETUP

We conduct experiments using the publicly available Librispeech dataset [25], where clean speech of 1172, 40 and 40 different speakers are utilized for training, development and test respectively. In our experiments, we select 500 and 1168 speakers from the training set to train the CVAE and the x-vector system, and the development and test set are combined together as the final test set to evaluate the performance of the systems to unseen speakers.

Table 1. Extraction performance for different RT settings.

| $T_{60}$ [s] | BSS Algorithm | Avg. Accuracy Rate [%] | | Avg. SDR improvement [dB] | | |
|---|---|---|---|---|---|---|
| | | Aug. a | Aug. b | Aug. a | Aug. b | Oracle |
| **0.16** | ILRMA | 98.61 | **99.17** | 7.94 | **8.02** | **8.19** |
| | MVAE | 98.61 | 98.47 | 7.64 | 7.68 | 7.88 |
| **0.36** | ILRMA | 98.33 | **98.33** | 6.05 | **6.12** | **6.24** |
| | MVAE | 97.78 | 97.64 | 5.68 | 5.70 | 5.84 |
| **0.61** | ILRMA | **95.28** | 94.58 | **5.22** | 5.18 | **5.31** |
| | MVAE | 94.58 | 93.89 | 4.93 | 4.85 | 5.07 |

### 4.1. Evaluation configuration

We evaluate the extraction performance of the proposed target source separation method using real RIRs from the multichannel impulse response database (MIRD) [26]. The RIRs are measured every 15 degrees, ranging from -90° to 90° around a linear microphone array with three RT settings of 0.16 s, 0.36 s, and 0.61 s. The 8 cm spacing configuration with a one-meter microphone-source distance is chosen for each RT setting, and only the middle two microphones are used in our experiments. Evaluation mixtures are created by convolving RIRs with two speech signals, and all possible configurations of incident angles are considered as long as the corresponding DOA interval lies in the range of 15° to 120°. Both the durations of the evaluation mixtures and the enrollment utterances are set to be 30 s, and the enrollment utterances are different from those used to generate evaluation mixtures. In total, we generate 360 mixtures with an initial signal-to-interference-ratio (SIR) selected from [-5, 0, 5] dB under each RT, which gives us a total number of 2160 trials by choosing either source as the target signal.

The basis number of NMF in the ILRMA algorithm is empirically set to be 2, as suggested by the result of [12], and the NMF parameters are randomly initialized. For the MVAE algorithm, we adopt the initialization strategy where the demixing matrices are warm restarted using the ILRMA run for 30 iterations as suggested in [13]. Both BSS algorithms are processed under a 16 Hz sampling rate, a 64 ms window length and a window overlap of 75%. Exemplary audio samples are available online at https://github.com/annie-gu/MVAEBasedBSE.

### 4.2. Results and discussion

We evaluate the extraction performance in terms of the extraction accuracy and the scale-invariant SDR, where the SDRs are calculated using the BSS_EVAL toolbox [27].

Fig. 2 (a) shows the accuracy rate of speaker selection for varied oracle SDRs and (b) presents the distribution of extracted utterances in terms of SDR improvements (SDRi). The oracle SDRs are obtained by finding the output permutation that maximizes the output SDR, and SDRis represent the SDR differences between extracted speech and mixed signals observed at the reference microphone.

We compare the extraction performance of x-vector trained with and without data augmentation (denoted as Without Aug.). Both Aug.a and Aug.b improve the accuracy rate in investigated SDR ranges and reduce the proportion of low SDR improvements (SDRi < 5 dB) caused by selection error. Note that although the augmented data in Aug.a is considered to be more consistent with the separated signals, Aug.b achieves comparatively similar results, which confirms the effectiveness of the generic augmentation scheme applied to the training of x-vector system.

From Fig. 1 (b), the capability of MVAE generalizing to unseen speakers is validated. However, unlike the 4-speaker case shown in [13], MVAE has no advantage over the more simplified rule-based ILRMA based system. The ILRMA system even produces more utterances of higher quality (SDRi > 5 dB) than the MVAE system.

Table 1 provides detailed results for different RT scenarios. Column indexed with 'Oracle' presents the averaged SDRi from oracle target speaker selection, which forms the performance upper-bound of the x-vector based speech extraction method. It can be seen that the ILRMA based system performs better than the MVAE based system. On the other hand, Aug.b based system performs slightly worse when reverberation is high, yet the differences between the two augmentation strategies are trivial, suggesting that the feature of the residual interference in the separated signals can be well characterized by the babble noise used in Aug.b.

### 5. CONCLUSION

In this paper, we investigate a sequential target speech extraction system by combining ILRMA or MVAE algorithm with an x-vector based speaker recognition module. The generalization of MVAE to unseen speakers is evaluated and compared with ILRMA. A dedicated data augmentation with BSS processed signals and a generic data augmentation with only babble noise and reverberation are utilized to train the x-vector system. From the experimental results it can be seen that MVAE trained with significantly more speakers can be generalized to unseen speakers, but it has no advantage over the more efficient ILRMA based system. Considering the overall performance and the computational efficiency, the ILRMA based BSS algorithm followed by the SR module trained with generic data augmentation is preferred.